\documentclass[10pt,twocolumn, nofootinbib]{revtex4-1}
\usepackage{amssymb,amsmath}
\usepackage{graphicx}
\usepackage{color}
\usepackage{hyperref}
\usepackage{listings}
\usepackage{natbib}
\begin{document}

\title{\Large Temporal Solitons in Microresonators driven by Optical Pulses}
\author{\textbf{Ewelina Obrzud$^{1,2}$, Steve Lecomte$^{1}$, Tobias Herr$^{1,*}$}}
\vspace{1cm}
\affiliation{\mbox{$^{1}$Swiss Center for Electronics and Microtechnology (CSEM), Time and Frequency Section, Neuch\^atel, Switzerland}
\\\mbox{$^{2}$Geneva Observatory, University of Geneva, Geneva, Switzerland}
\\ \\$^*$Tobias.Herr@csem.ch}

\begin{abstract}
Continuous-wave laser driven Kerr-nonlinear, optical microresonators have enabled a variety of novel applications and phenomena including the generation of optical frequency combs, ultra-low noise microwaves, as well as, ultra-short optical pulses. In this work we break with the paradigm of the continuous-wave optical drive and use instead periodic, pico-second optical pulses. We observe the deterministic generation of stable femtosecond dissipative cavity solitons on-top of the resonantly enhanced driving pulse. Surprisingly, the soliton pulse locks to the driving pulse enabling direct all-optical control of both the soliton's repetition rate and carrier-envelope offset frequency without the need for any actuation on the microresonator. When compared to both continuous-wave driven microresonators and non-resonant pulsed supercontinuum generation, this new approach is substantially more efficient and can yield broadband frequency combs at femto-Joule driving pulse energies and average laser powers significantly below the parametric threshold power of continuous-wave driven microresonators. The presented results bridge the fields of continuous-wave driven resonant and pulse-driven non-resonant nonlinear optics. They enables micro-photonic pulse compression, ultra-efficient low noise frequency comb and resonant supercontinuum generation for applications including optical data transfer and optical spectroscopy. From a scientific perspective the results open a new horizon for nonlinear photonics driven by temporally and spectrally structured light.

\end{abstract}

\maketitle

\section{Introduction}

Continuous-wave (CW) laser driven, Kerr-nonlinear optical microresonators\cite{vahala2003,braginsky1989,grudinin2006,armani2003,razzari2010,levy2010,agha2007,hausmann2014,delhaye2013,jung2013} have emerged as versatile platforms for nonlinear optics. They give rise to efficient parametric frequency conversion\cite{kippenberg2004b,savchenkov2004,guo2016}, stimulated Brillouin\cite{grudinin2009,lee2012} and Raman scattering\cite{spillane2002, kippenberg2004a,grudinin2007}. In particular, microresonator based frequency combs with tens of Gigahertz mode spacing\cite{delhaye2007,kippenberg2011,savchenkov2008,papp2011,okawachi2011,li2012,griffith2015,hausmann2014} from visible\cite{miller2014,savchenkov2011,wang2016b} to mid-infrared wavelength\cite{griffith2015,wang2013,yu2016} and over wavelengths intervals that enable self-referencing\cite{okawachi2011,delhaye2011,jost2015,delhaye2016} have attracted significant attention due to their high technological potential. Already now, such frequency combs have found application in coherent optical data transfer\cite{pfeifle2014},  arbitrary optical waveform generation\cite{ferdous2011}, ultra-low noise electronic signal generation\cite{savchenkov2008,li2013,liang2015} and also hold promise for astronomical spectrometer calibration\cite{li2008,steinmetz2008,yi2016}. 
Of particular importance is the generation of temporal dissipative Kerr-cavity solitons (DKS)\cite{herr2014a,yi2015,brasch2016,wang2016a,webb2016,joshi2016,cole2016,lobanov2016}, ultra-short bright pulses of light that propagate indefinitely and without changing their shape inside the microresonator. Such DKS provide a reliable way of achieving a smooth spectral envelope and low noise frequency combs\cite{herr2014a,herr2012}. Owing to these properties, such soliton frequency combs have for instance enabled chip-scale dual comb spectrometers\cite{suh2016,dutt2016}, 50~Tbit/s optical data transmission\cite{marin2016} and broadband, low noise comb generation for self-referencing\cite{brasch2016,jost2015}.  Essential for the existence of (bright) DKS are the optical nonlinearity and the material's anomalous group velocity dispersion (GVD) that, by balancing each other, prevent temporal spreading of the soliton. In addition, the inevitable resonator losses are compensated by the CW driving laser or ``holding beam'' that via the Kerr-nonlinear parametric gain constantly provides energy to the soliton.  While their spatial analog has been known to scientists for more than a decade\cite{barland2002}, temporal DKS, a manifestation of the intricate yet stable interplay between dispersion, nonlinearity, parametric gain and loss, have only recently (prior to their observation in microresonator) been observed in long fiber-ring cavities\cite{leo2010}. Here, the fiber-ring cavity was driven by a CW holding beam and individual picosecond pulses have been used to excite DKS. Moreover, it has been shown in fiber-ring cavities that phase modulation of the CW holding beam on time scales much faster than the resonator roundtrip time can be used for ‘writing’ and ‘erasing’ of DKS at arbitrary times inside the cavity\cite{jang2015b} and for trapping solitons into specific (co-propagating) timeslots\cite{jang2015a}. Such methods are not directly applicable to microresonators as the resonator roundtrip time is much faster and the soliton pulse duration much shorter in comparison to fiber-ring cavities. Indeed, the generation of DKS in microresonators is to date challenging. First, a precise and rapid control (comparable or faster than the resonator's thermal time constant) of the relative CW laser to resonator detuning is required in order to achieve stable soliton operation\cite{herr2014a}.  Second, due to the spontaneous formation of the DKS from an uncontrollable breathing soliton state\cite{herr2014a,matsko2012,matsko2013,lucas2016}, the number as well as the relative separation in time between solitons are random and cannot deterministically be controlled. While to date operation of nonlinear microresonators does generally rely on a CW driving laser, it has been shown that phase and amplitude modulation of the CW laser can overcome some of the challenges associated with the formation of DKS\cite{lobanov2016,taheri2015}. These approaches are similar to experimental demonstrations where bi-chromatic driving\cite{strekalov2009} or parametric seeding\cite{papp2013a} were used to control the dynamics of four-wave mixing or the generation of ``platicons'' in the normal dispersion regime\cite{lobanov2015} via intensity modulation of the CW laser. In all cases however, a CW driving beam remains an integral component of the system. Such CW operation is in contrast to non-resonant schemes in nonlinear optics where, e.g. in supercontinuum generation\cite{dudley2006}, in order to benefit from nonlinear optical effects ultra-short, high peak power pulses are used (the peak power enhancement corresponds to the inverse pulse duty cycle).
In this work we deviate from the concept of a CW driving beam for micoresonators. Instead, we explore resonantly enhanced pulsed driving of a microresonator for nonlinear optics and, in particular, the generation of DKS. This does not only correspond to a novel scheme of resonant supercontinuum generation, but could also allow for unprecedentedly efficient, microresonator-based ultra-short soliton pulse and frequency comb generation. In a pulsed driving configuration, a periodic train of driving pulses would need to replace the CW holding beam, so that inside the cavity a resonantly enhanced pulse co-propagates with the soliton pulse (Fig.\ref{fig1}a,b). Based on the current understanding, this would however require fine-tuning of the driving pulse repetition rate to avoid that the driving pulse and the soliton drift apart\cite{malinowski2016}. Considering a driving pulse duration of 1 ps and a driving pulse repetition rate of 10 GHz, it can be expected that the soliton and the driving pulse would drift apart (resulting in annihilation of the soliton) within less than 10 ms if the driving pulse repetition rate differs only by 1 Hz from the soliton's natural pulse repetition rate (defined by the free-spectral range (FSR) of the microresonator). Stabilizing the microresonator's FSR to the 1 Hz level would amongst other aspects imply a temperature stability of better than 10 $\mu$K, despite the effect of comparably strong laser induced heating. Hence, stable DKS generation seems exceedingly challenging in an experimental system.

\section{Results}

Below, we investigate for the first time theoretically and experimentally the dynamics of nonlinear optical microresonators driven by periodic optical pico-second pulses, whose corresponding optical modes are matched to the modes of the resonator. We show that femtosecond DKS can form inside the cavity on-top of the resonantly enhanced external driving pulses. Surprisingly, DKS formation does not require the external pulse repetition rate to exactly match the FSR of the resonator. Instead, the repetition rate of the external laser can even be tuned around this value without destroying the generated soliton. The measurements reveal that the soliton stays tightly locked to the driving pulses and adapts adiabatically to the externally imposed pulse repetition rate. This remarkable behavior is in agreement with numerical simulations revealing the underlying plasticity of the soliton that by shifting its center wavelength adapts to the externally imposed pulse repetition rate. We moreover demonstrate that pulsed driving can significantly lower the average threshold power of DKS formation to below the parametric threshold power\cite{kippenberg2004b} (and even below the thermal bistability power\cite{carmon2004}) of CW driven systems. An advantageous side effect of the highly efficient and ‘targeted’ pulsed driving is that, in contrast to CW driven systems, the laser can be slowly (i.e. manually) tuned into the soliton state without the need for rapid and complex actuation on the driving laser or the resonator. Finally, we note that for the experiments a novel fiber-based microresonator in Fabry-P\'erot\cite{obrzud2016} geometry is used, constituting the first demonstration of temporal dissipative solitons in a standing wave resonator.

\subsection{Microresonator and Experimental Setup}

	\begin{figure*}[]
	\centering
	\includegraphics[width=0.75\textwidth]{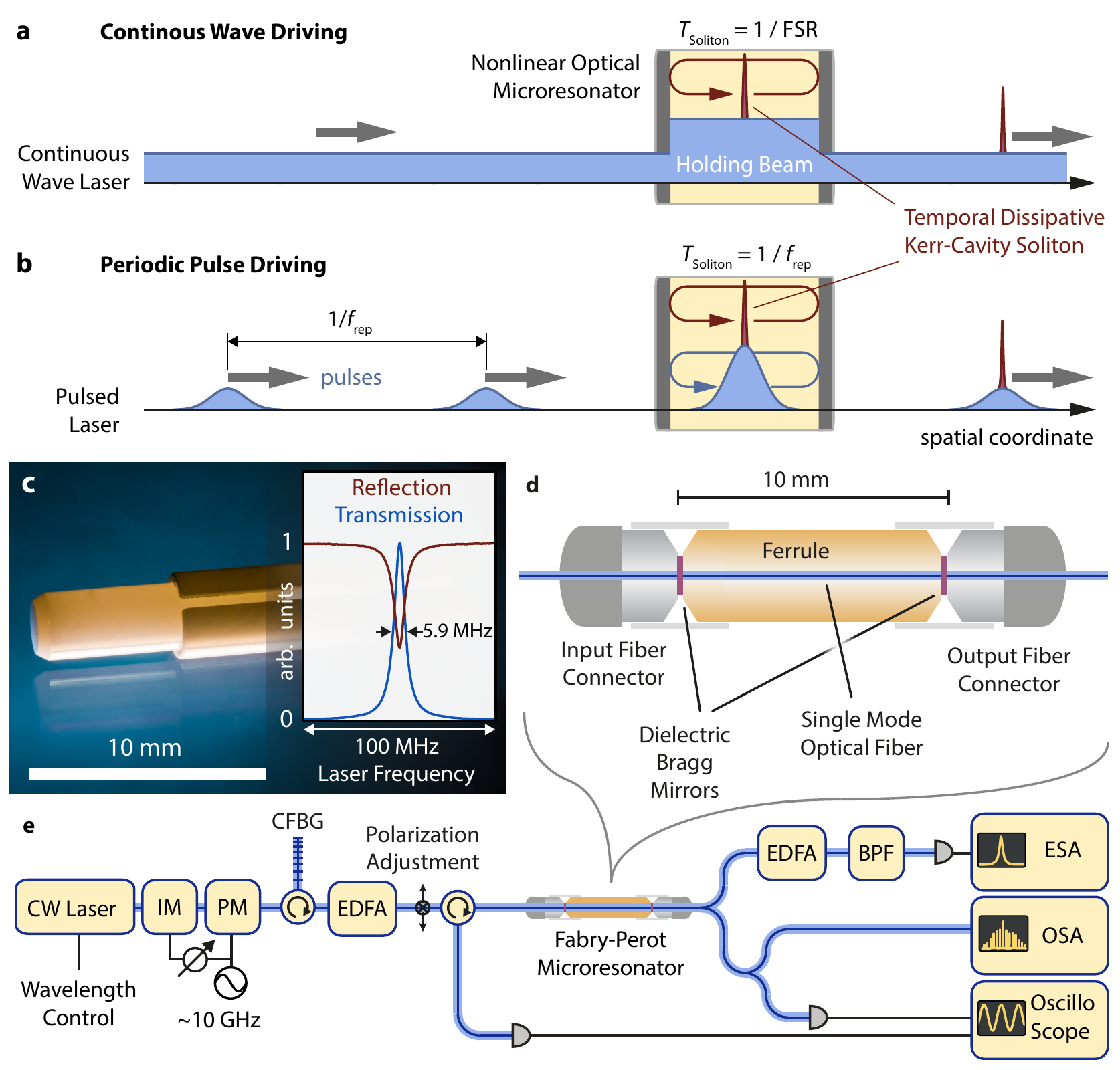}
	\caption{\small \textbf{Experimental scheme and setup.} (a) Continuous-wave (CW) driven microresonator: A temporal dissipative soliton can form inside the resonator and propagate with a roundtrip time defined by the resonator's inverse free-spectral range (FSR) while being supported by the resonantly enhanced CW background. (b) Pulse driven microresonator: Periodic pulses can resonantly build up in the resonator when their corresponding optical modes coincide with the resonance frequencies. Stable solitons can only form if the inverse driving pulse repetition rate $1/f_\mathrm{{rep}}$ and the soliton roundtrip time match precisely. (c) Photograph of a fiber-based Fabry-P\'erot microresonator. The inset shows the measured reflection and transmission when scanning a low-power CW laser across the resonance frequency. (d) Schematic of the microresonator and its integration into the setup. (e) Experimental setup comprising a pico-second periodic laser pulse generator based on electro-optic intensity and phase modulators (IM/PM) driven by a 10 GHz signal generator. A chirped fiber-Bragg grating (CFBG) is used for chirp-compensation and pulse formation. EDFA: erbium doped fiber amplifier, BPF: bandpass filter, ESA: electronic spectrum analyzer, OSA: optical spectrum analyzer.}
	\label{fig1}
	\end{figure*}

In the experiment we use a novel single-mode optical fiber-based Fabry-P\'erot microresonator\cite{obrzud2016} as shown in Fig.\ref{fig1}c. This microresonator consist of a mm-scale optical fiber whose end facets have been coated with highly reflective, zero group delay dielectric Bragg-mirrors. The resonator has a FSR of 9.77 GHz, a resonance width of 5.9 MHz and a linear coupling efficiency of 60\%. While the Bragg-mirror would in principle allow for engineering of the resonator's group-velocity dispersion (GVD) e.g. for operation in the visible wavelength regime or for dark pulse generation\cite{liang2014, xue2015}, this is not required in the present case, where the optical fiber already naturally provides an anomalous GVD of $\beta_2= -20\, \mathrm{ps^2/km}$ (at the driving center wavelength of 1559 nm) as required for the existence of bright solitons. The anomalous dispersion implies that the resonator's FSR increases with optical frequency $\nu$. The Kerr-nonlinearity and the effective mode area of the optical fused silica fiber are $n_\mathrm{2}=0.9 \times 10^{-20}\, \mathrm{m^2W^{-1}}$ and $A_\mathrm{eff}=85\, \mathrm{\mu m^2}$. This resonator design allows for a high-Q microresonator with a FSR low enough to be matched by the available driving laser source. For reasons of mechanical robustness and convenient interfacing with optical fiber, the resonator fiber is mounted inside a fiber optic ferrule whose diameter matches standard FC/PC fiber connectors. As opposed to travelling wave resonators, which require a bus waveguide, a prism or a tapered optical fiber for coupling, the present system is inherently fiber-coupled (Fig.\ref{fig1}d). The transmitted spectrum gives direct access to the intracavity soliton field and is equivalent to a drop port that efficiently suppresses uncoupled driving light during operation\cite{wang2016a}. While it is the first time that generation of DKS is attempted in a Fabry-P\'erot microresonator, this resonator geometry is formally equivalent to ring-type microresonators. We note that a similar nonlinear resonator, albeit with lower finesse and smaller FSR has been used for CW driven Brillouin-enhanced hyper-parametric generation\cite{braje2009}.

Figure \ref{fig1}e shows the experimental setup. While in principle the driving laser can be based on any pulsed laser technology, we use for experimental exploration an electro-optic modulation (EOM) based pico-second pulse generator\cite{kobayashi1988}. A 1559 nm CW fiber laser is strongly chirped using an EOM phase modulator (driven by a tunable 9.77 GHz microwave source) and compressed into picosecond pulses via linear propagation in a chirped fiber-Bragg grating with a group delay dispersion (GDD) of 10 ps/nm. Prior to phase modulation, an intensity modulator is used to carve out the modulation half-period with the correct sign of chirp. Such EOM based pulse generators allow for straightforward control of both pulse repetition rate and carrier-envelope offset frequency (equivalent to the optical frequency of the CW laser). After chirp compensation in the CFBG the pulses have a pulse duration of 2.4 ps and are characterized by an almost flat-top spectrum. Next, the pulses are amplified in an erbium-doped fiber amplifier up to 1.5 W of average power. Prior to being coupled to the resonator, the amplified pulses propagate through approximately 10 m of optical fiber resulting in moderate pulse shortening to 2.1 ps and formation of weak side pulses (peak power 13\% of main pulse, separation from main pulse approx. 2 ps). This manifests itself in the generation of ``spectral ears'' reducing the efficiency of the driving laser. While not critical for the present work, this could be improved by optimizing the delivery of the amplified pulses to the microresonator. Throughout this work we assume to effectively generate 30-50 optical lines (corresponding to the inverse pulse duty cycle) of approximately equal optical power that are spaced from each other in optical frequency by the modulation frequency.

\subsection{Experimental Generation of Solitons via Pulsed Driving.}

\begin{figure*}[]
	\centering
	\includegraphics[width=0.75\textwidth]{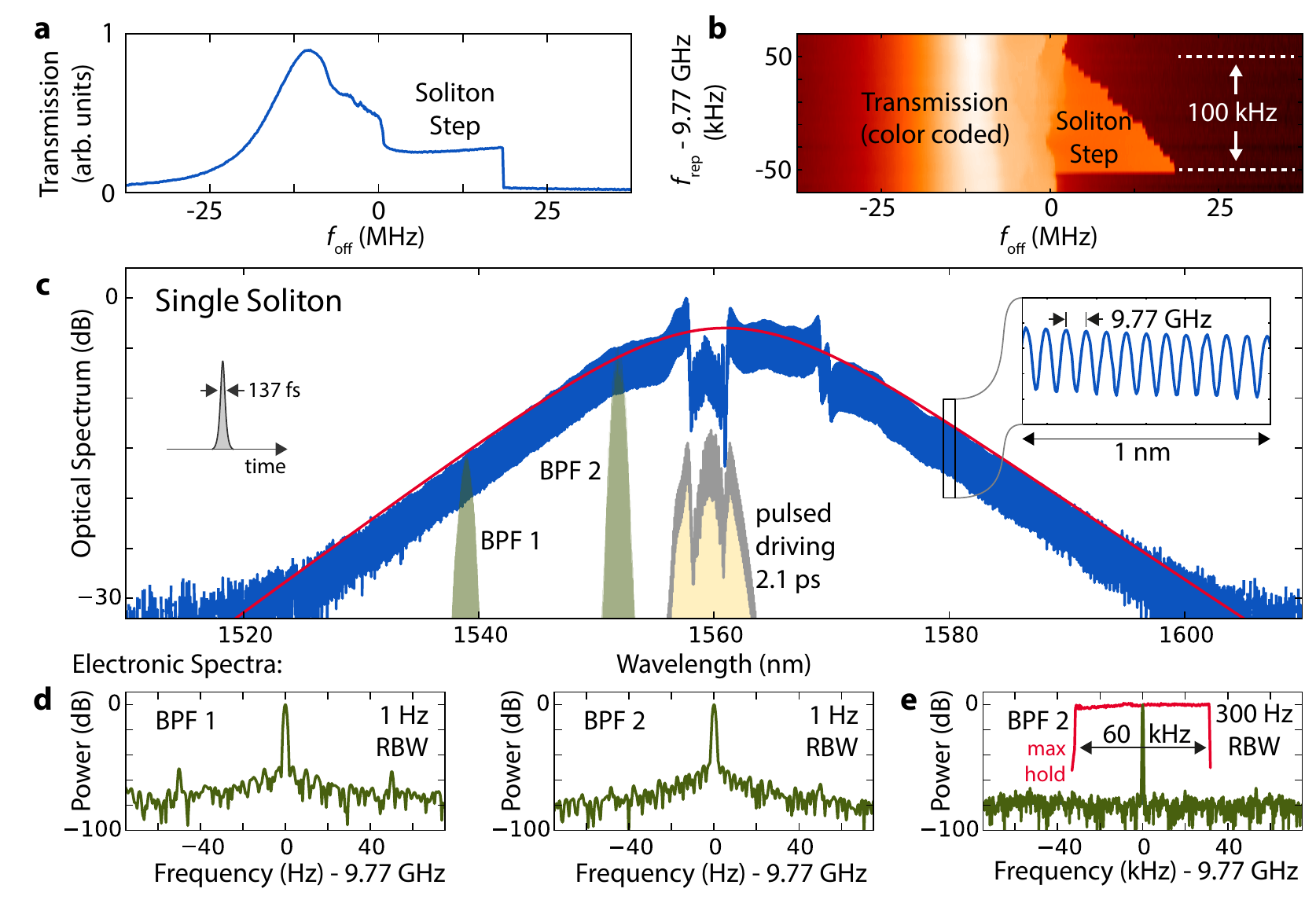}
	\caption{\small \textbf{Experimental soliton generation.} (a) Resonator transmission obtained when scanning the pulse offset frequency $f_\mathrm{off}$ across a resonance frequency for varying pulse repetition rate  $f_\mathrm{rep}$. (b) 2D contour plot showing the nonlinearly deformed resonance shape including ``soliton steps'' that appear within an 100~kHz wide interval of $f_\mathrm{rep}$ when scanning the pulse offset frequency $f_\mathrm{off}$ across a resonance frequency. (c) Optical spectrum of a single soliton obtained when driving with 2.1 ps pulses and a coupled power of maximally 100 mW. The sech$^2$-envelope fit (red) indicates a soliton pulse duration of 137 fs. The yellow/grey spectrum represents the spectrum of the driving pulses. The green spectra show the transmission spectra of the bandpass filters (BPF 1 and 2) used for intermode beatnote detection. (d) Intermode beatnotes recorded using BPF 1 and 2 in the setup shown in Fig.\ref{fig1}e. A resolution bandwidth (RBW) of 1 Hz is used. (e) Adiabatic change of the soliton's intermode beatnote (i.e. soliton repetition rate) by varying the external pulse repetition rate $f_\mathrm{rep}$ within an interval of 60 kHz (red is the max-hold trace of the green signal; BPF 2 is used). }
	\label{fig2}
	\end{figure*}

In a first experiment we attempt to generate solitons by precisely matching first the pulse repetition rate (defined by the modulation frequency) to the resonator's FSR and second the central mode of the pulsed driving laser (defined by the CW fiber laser in Fig.\ref{fig1}e) to a resonance of the microresonator. The central spectral mode of the driving pulses corresponds to the carrier-envelope offset frequency of the driving pulses. In the following we denote this quantity as $f_\mathrm{off}$ and measure it relative to an (arbitrary) optical frequency close to a resonance frequency. In this first experiment, the pulse duration is 2.1 ps and the coupled driving power approximately 100 mW. In order to identify a regime of potential soliton formation, we repeatedly scan $f_\mathrm{off}$ across a resonance (from blue to red detuning in analogy to the CW driven system\cite{herr2014a,carmon2004}) while at the same time slowly varying the microwave modulation frequency around the value of the FSR of 9.77 GHz. The resonator transmission, which is the cumulative resonance shape of all driven modes, shows characteristic ``step'' features. These are similar to the ones observed in CW driven systems where they are directly related to DKS formation\cite{herr2014a} (Fig.\ref{fig2}a). In contrast to CW driven systems, the observed transmission is essentially free of the random fluctuation of step length and height. Surprisingly, the step feature appears for a rather wide, 100 kHz spanning interval of driving pulse repetition rates $f_\mathrm{rep}$, suggesting that DKS formation in a pulsed system is unexpectedly robust against a mismatch between $f_\mathrm{rep}$ and the resonator's FSR.

Indeed, when tuning into the soliton step, we observe a spectrum of more than 1000 individual spectral lines (within 25 dB) and its envelope follows closely the sech$^{2}$-shape characteristic for DKS (Fig.\ref{fig2}c). This spectrum corresponds to a single soliton pulse of 137 fs duration circulating stably in the microresonator with a pulse repetition rate of 9.77 GHz. Distinct from CW driven systems is that there is no strong individual spectral component being orders of magnitude stronger than the soliton spectrum. Two spectral features can be distinguished: First, around 1570 nm an ``up-down'' feature, characteristic for avoided mode crossings\cite{herr2014b} is visible. While the intrinsically single mode resonator drastically reduces the number of mode crossings (compared to mostly multi-mode microresonators), the two polarization mode-families (non-degenerate due to stress induced birefringence) can weakly couple via linear scattering at the resonator facets. This coupling can give rise to avoided mode crossings. The second spectral feature, is composed of the spectral ``ears'' adjacent to the central portion of the driving laser spectrum. Those ``ears'' are already present in the driving laser spectrum prior to its coupling to the microresonator and are due to the Kerr-nonlinearity of the optical fiber transporting the pulsed light from the EDFA to the microresonator. Only the central in-phase portion of the spectrum effectively contributes to driving the soliton (as can be seen from the relative power reduction of this part of the driving spectrum).

Once tuned into the soliton state, the soliton remains stable until the laser is tuned out of the resonance without requiring any feedback loop. Figure \ref{fig2}d shows the electronic intermode beatnotes detected after optical bandpass filtering and amplification of the soliton spectrum (according to Fig.\ref{fig1}e and Fig.\ref{fig2}c). The bandpass filtering limits the detection to spectral components of the actual soliton pulses and not the driving pulses, the amplification ensures a good signal-to-noise ratio. The detected beatnotes precisely correspond to the repetition rate of the driving laser (microwave synthesizer and analyzer were cross-referenced), are perfectly stable and show 1 Hz resolution bandwidth limited signals without any sidebands or sign of noise anywhere from DC to the carrier frequency (the weak sidebands visible in Fig.\ref{fig2}d (left) at 50 Hz offset frequency are induced by the electrical power lines). 

	\begin{figure*}[t]
	\centering
	\includegraphics[width=0.75\textwidth]{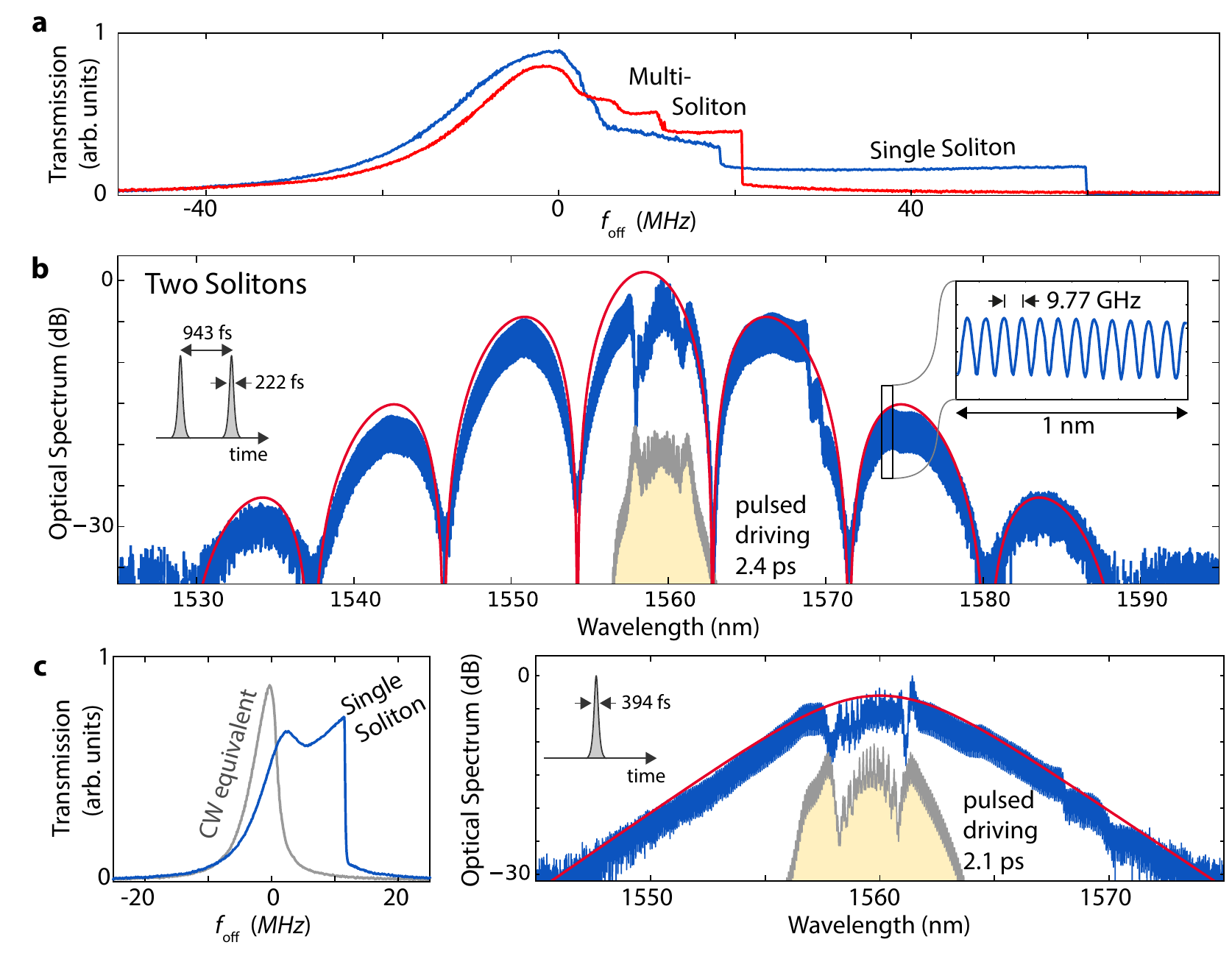}
	\caption{\small \textbf{Multi-solitons and low-power driving.} (a) Comparison of single and multiple (two) solitons transmission curves. (b) Optical spectrum of two solitons obtained when driving with 2.4 ps pulses and a coupled power of maximally 45 mW. The modulated sech$^2$-envelope fit (red) takes the interference of the soliton into account and yields a soliton pulse duration and separation of 222 fs and 943 fs, respectively. The yellow/grey spectrum indicates the spectrum of the driving pulses. (c) Left: Comparison of transmission curves for driving with a CW laser (grey) and 2.1 ps/9.8 GHz pulses (blue) with 3 mW of coupled average power. Right: Experimentally generated spectrum obtained when tuning into the single soliton ‘step’ shown in the left panel. Based on the sech$^2$-envelope fit (red) the single soliton pulse duration is estimated to be with 394 fs. The used strongly under-coupled resonator has a resonance width of 2.5 MHz and a FSR of 9.8 GHz.}
	\label{fig3}
	\end{figure*}

Next, we investigate the robustness of the solitons against variation of the driving laser's pulse repetition rate $f_\mathrm{rep}$. When varying the driving pulse repetition rate $f_\mathrm{rep}$, the soliton pulse repetition rate follows adiabatically (i.e. without annihilation of the soliton) over a frequency interval spanning 60 kHz (Fig.\ref{fig2}e). As a result, the soliton's repetition rate can be all-optically controlled and stabilized, in contrast to CW driven systems, where it is impacted by resonator fluctuation, fluctuation of laser detuning or laser power and the Raman soliton self-frequency shift\cite{bao2016, karpov2016}. The observed experimental behavior suggests that the soliton locks to the driving pulse and adapts to the externally imposed repetition rate. We will identify and discuss the underlying physical mechanism later on.
Meanwhile, it is interesting to compare the interval of 60 kHz over which locking occurs to the change of the FSR occurring when the resonator's temperature increases due to laser induced heating. A direct measure of this heating effect is the resonance frequency shift of approximately 10-100 MHz, occurring when the pulsed laser is tuned into soliton operation (cf. Fig.\ref{fig2}c). This corresponds to a change of the FSR of only a few kilohertz and is much smaller than the tolerated frequency mismatch. This explains the robust operation of the resonator under pulsed driving even when the resonator is not actively stabilized. Moreover, this tolerance of the soliton against temperature variation also enables ``thermal locking''\cite{carmon2007}, a self-stabilizing effect that locks the resonator's resonances to the driving laser as long as the intracavity power has a negative slope as a function of optical laser frequency (here: $f_{off}$). Owing to its mechanical stability, the coupled resonator does not require any acoustic or vibration damping, despite its direct connection to other fiber optical components.

Typically, we only observe step features in the transmission that correspond to single soliton states. This enables deterministic and reliable generation of single DKS states, which is challenging to achieve in CW driven systems, yet highly desirable, as the resulting optical spectrum is characterized by a smooth, unmodulated sech$^2$-envelope (Fig.\ref{fig2}c). However, via (de-)tuning e.g. the driving pulse repetition rate $f_\mathrm{rep}$, the driving pulses can be arranged to also deterministically result in multi-soliton states as evidenced by higher ``step height'' (Fig.\ref{fig3}a). The corresponding optical spectra are characterized by a modulated envelope as shown in Fig.\ref{fig3}b, where the spectral modulation corresponds to the inverse separation of the soliton pulses in time. Remarkably, identical multi-soliton states can be reproduced by using identical driving pulse parameters. The stability of the spectral envelope of multi-soliton states over long time scales ($>$ 30 minutes) also agrees well with the solitons being not only loosely attached to but indeed tightly locked to a specific relative temporal position of the driving pulse. 

	\begin{figure*}[t]
	\centering
	\includegraphics[width=0.75\textwidth]{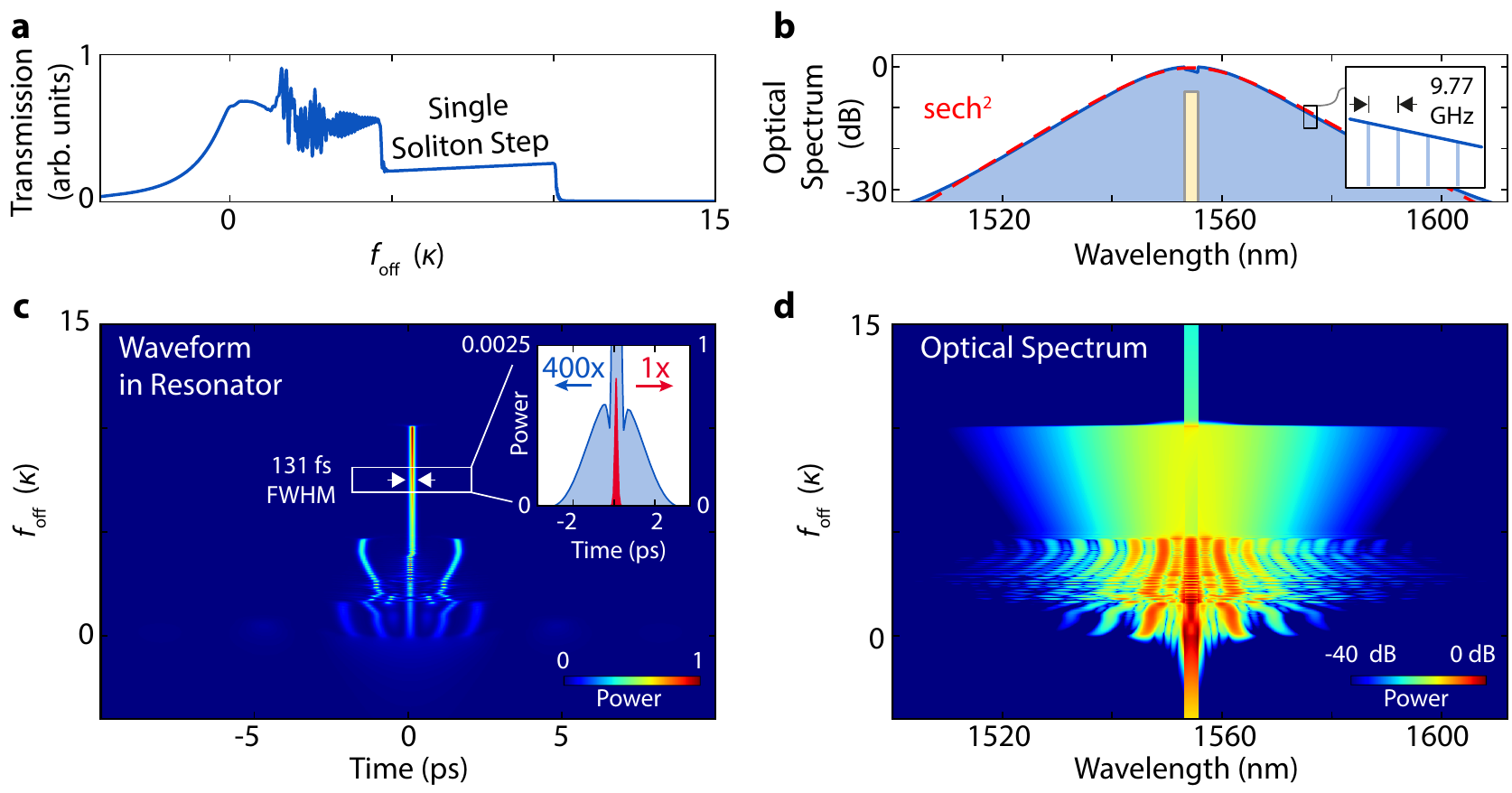}
	\caption{\small \textbf{Numerical simulation of soliton formation in a pulse-driven microresonator.} (a) Simulated resonator transmission when a driving pulse train with $f_\mathrm{rep} = \mathrm{FSR}$ is coupled to a microresonator and the pulse offset frequency $f_\mathrm{off}$ is scanned across a resonance. A characteristic ‘step-feature’ resulting from the formation of a single soliton is visible. The driving pulses are assumed to be composed of 31 spectral components with equal phase and power. The simulated microresonator is characterized by a resonance width of $\kappa/(2\pi)$ =7 MHz, a FSR of 9.77 GHz, a Kerr-nonlinearity of  $n_\mathrm{2}=0.9 \times 10^{-20}\,  \mathrm{m^2W^{-1}}$, an effective mode area of $A_\mathrm{eff}=85\, \mathrm{\mu m^2}$and a group velocity dispersion of $\beta_2=-20\, \mathrm{ps^2/km}$.  (b) Optical spectrum of the single soliton at the pulse offset frequency of $f_\mathrm{off}/\kappa$=7.5 (measured with respect to the cold cavity's resonance frequency) showing the sech$^2$-envelope that is characteristic for a soliton. The yellow area represents the spectral content of the driving pulses. (c) Evolution of the intracavity waveform as $f_\mathrm{off}$ is tuned across the resonance (only a 10 ps wide window, co-moving with the resonantly enhanced driving pulse is shown). (d) Same as (c) but the optical spectrum is shown.}
	\label{fig4}
	\end{figure*}

Pulsed driving allows for a significant reduction of the required average driving laser power when compared to a CW driven system (scaling with the pulse duty cycle). Consequently, the associated laser induced heating and thermal resonance shifts are reduced. In stark contrast to CW laser driven microresonator based DKS generation where rapid laser wavelength tuning ramps, short driving power drops or rapid actuation on the microresonator are required to stably enter the soliton state\cite{herr2014a,brasch2016,webb2016,joshi2016}, it is here possible to slowly (and manually) tune the pulsed laser into the soliton state by increasing the wavelength of the driving modes until the step feature is reached. Figure~\ref{fig3}c illustrates the reduction of required driving power by comparing both the CW and pulsed driving. In the pulsed case, single soliton generation is already possible when driving the resonator (linewidth 2.5 MHz, FSR 9.8 GHz) with 2.1 ps pulses and 3 mW of maximally coupled average power corresponding to pulse energies of approximately 300 fJ. The required average power is below the parametric threshold power\cite{kippenberg2004b} and even below the thermal bistability power\cite{carmon2004}.

	\begin{figure*}[t]
	\centering
	\includegraphics[width=0.75\textwidth]{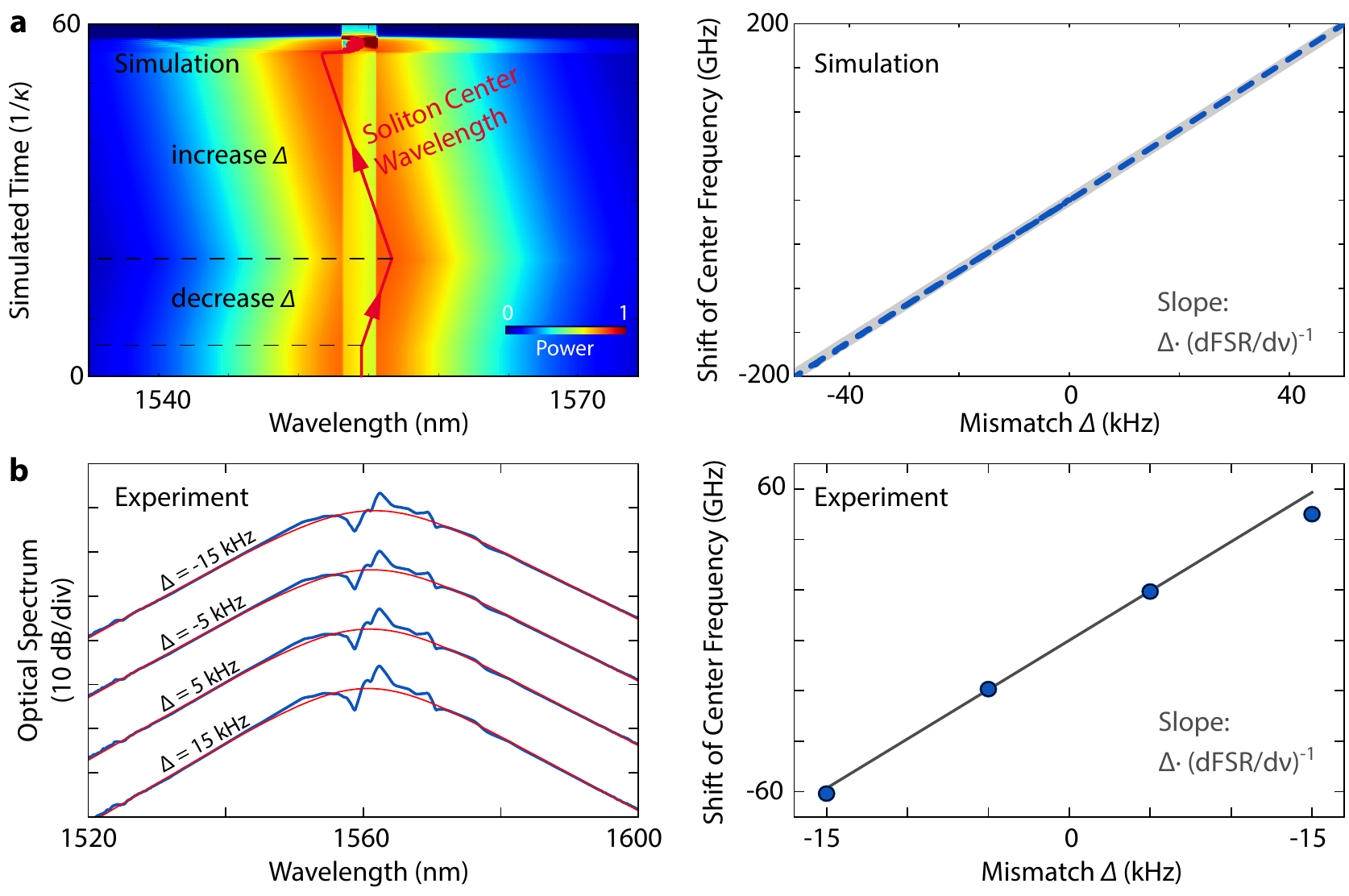}
	\caption{\small \textbf{Soliton-to-pulse-locking.} (a) Left: Simulated evolution of the soliton's central wavelength when, for a fixed value of $f_\mathrm{off}/\kappa$=7.5 (cf. Fig.\ref{fig4}), the pulse repetition rate is varied so that the mismatch parameter $\Delta$ becomes first negative, then again zero and positive until the soliton decays. Right: Simulated shift of the soliton's center frequency as a function of the applied mismatch parameter $\Delta$. The grey line corresponds to a linear function with a slope of $\Delta \cdot \left(\mathrm{dFSR}/\mathrm{d}\nu \right)^{-1} $ as discussed in the main text. (b) Left: Experimentally generated, single soliton spectra (1 nm resolution and vertically offset for better visibility) for four different values of the mismatch parameter $\Delta$ (relatively measured). The sech$^2$-envelopes fitted to the spectral wings (red) allow deriving the soliton's center wavelength/frequency. Right: Derived shift of soliton center frequency as a function of the of applied mismatch parameter $\Delta$. The grey line corresponds to a linear function with a slope of $\Delta \cdot \left(\mathrm{dFSR}/\mathrm{d}\nu \right)^{-1} $ as discussed in the main text. }
	\label{fig5}
	\end{figure*}

\subsection{Numerical simulations}

In order to interpret the experimental results and to corroborate the hypothesis of the soliton locking to the driving pulse and its repetition rate, we perform numerical simulations of the experimental resonator system that are based on the nonlinear coupled mode equations\cite{herr2014a,lobanov2016,chembo2010}. The nonlinear coupled mode equations describe, in the frequency domain, the nonlinear interactions of the optical modes that are excited inside the resonator through external driving and nonlinear optical frequency conversion. The pulsed driving can be described by a set of in-phase driving modes separated by the FSR of the microresonator with intensities defined by the assumed temporal shape of the driving pulse. Here, for simplicity, we assume that all driving modes have the same intensity. In agreement with the experiment, the simulated resonator transmission is distorted by nonlinear optical effects and displays a step feature towards longer wavelengths (Fig.\ref{fig4}a). Coinciding with the step feature is a broadband optical spectrum well described by a sech$^2$-envelope (Fig.\ref{fig4}b). Figure~\ref{fig4}c,d show the simulated temporal and spectral evolution of the intra-cavity fields. Indeed, an ultra-short soliton pulse forms on-top of the much broader and approximately 400 times lower peak-power, intracavity driving pulse.  Interestingly, the solitons can be offset from the driving pulse center. Moreover, the simulations indicate that the soliton pulse stays at the same position in the temporal reference frame of the underlying driving pulse. The simulated generation of soliton is possible over a wide range of driving powers, where higher levels of power result in shorter duration DKS and broader optical spectra. Moreover, multi-soliton states can be generated when the driving pulses are broad enough to support two solitons, or when temporally structured light, i.e. pulses with complex envelope or multiple pulses, is used for driving. The Numerical simulations also confirm that the required power for single soliton generation scales approximately with the inverse number of in-phase driving modes (i.e. with the duty cycle of the driving pulse train).

\subsection{Soliton-to-Pulse-Locking}

As discussed in the beginning, the stable generation of DKS in an experimental pulsed driving configuration is surprising, as it is expected to be highly sensitive to minute differences between the resonator's FSR and driving pulse repetition rate $f_\mathrm{rep}$. However, the experimental observations (cf. Fig. \ref{fig2}, \ref{fig3}) suggest that the soliton pulse matches the externally imposed pulse repetition rate $f_\mathrm{rep}$ and locks to the driving pulse inside the resonator. In order to identify the underlying mechanism, we first define a mismatch parameter $\Delta=f_\mathrm{rep}-\mathrm{FSR}$, between the driving pulses' repetition rate $f_\mathrm{rep}$ and the free-spectral range at the drive's center wavelength. In order to study the locking mechanism in the absence of thermal effects that may impact an experimental study, we first perform a numerical simulation (similar as in Fig.~\ref{fig4} above but extended by the mismatch parameter $\Delta$). Figure \ref{fig5}a shows the simulation over a time corresponding to 60 cavity lifetimes. This simulation is seeded with the soliton generated in the previously described simulation (Fig.\ref{fig4}) at $f_\mathrm{off}/\kappa=7.5$. Then, in contrast to Fig.\ref{fig4}, $f_\mathrm{off}$ is kept constant and the mismatch parameter $\Delta$ is varied, first taking negative then positive values until the soliton decays. It can be seen in Fig.\ref{fig5}a (left) that the center wavelength of the soliton shifts as a function of the mismatch parameter $\Delta$ (the center wavelength/frequency is obtained via fitting the wings of the sech$^2$-envelope). As the resonator's FSR depends on the optical frequency, the shift of soliton's center wavelength entails a change of the soliton's repetition rate. For the present (experimental and simulated) microresonator this increase of FSR (and hence the natural soliton pulse repetition rate) can be approximated by $\mathrm{d}\mathrm{FSR}/\mathrm{d}\nu=256\, \mathrm{Hz}/\mathrm{GHz}$ at the driving center wavelength of 1559 nm. In principle, in order to obtain matching soliton and driving pulse repetition rates, the soliton would need to shift its center frequency by $\Delta \cdot \left(\mathrm{dFSR}/\mathrm{d}\nu \right)^{-1} $. Figure \ref{fig5}a (right), which is based on data extracted from Fig.\ref{fig5}a (left), confirms exactly this and hence reveals the physics of the soliton-to-driving pulse locking mechanism. This shift of soliton center wavelength, a remarkable mechanism of soliton self-organization and plasticity, is phenomenologically similar to $\chi^{(2)}$-nonlinear, synchronously pumped optical parametric oscillators, where the intracavity pulse shifts its center wavelength in order to match the externally imposed repetition rate of the pump laser\cite{kafka1995}. Finally, in order to experimentally confirm the identified locking mechanism, we experimentally generate optical soliton spectra for four different settings of the driving pulse repetition rate $f_\mathrm{rep}$(or respectively, the mismatch parameter $\Delta$) as shown in Fig.\ref{fig5}b (left). In order to reduce potential thermal effects or drifts, we record the spectra at high scan speed with low resolution. Via sech$^2$-envelope fitting in the wings of the optical spectrum, we determine the shift of the soliton's center wavelength and compare it in Fig.\ref{fig5}b (right) to the prediction based on the above considerations. Indeed, the experiment confirms the simulations and further corroborates the identified locking mechanism.

\section{Discussion}

In summary, we have investigated experimentally and theoretically pulsed driving of a microresonator for nonlinear optics and in particular for generation of temporal dissipative Kerr-cavity solitons (DKS). This approach is different from previous work that always required a CW driving or holding beam. When the optical frequency components of the driving pulse train are approximately matched with the modes of the resonatorwe, we observe the generation of stable DKS with femtosecond pulse duration on-top of the resonantly enhanced driving pulse. The required average driving laser power for generating the solitons is drastically reduced when compared to the conventional CW driving case. This is similar to dark pulse generation where the overlap of bright portions of the intracavity waveform with the driving light is optimized\cite{liang2014, xue2015}. Reducing the required driving power also reduces absorptive heating induced thermal effects in the resonator, which, in consequence, allows to slowly tune into the soliton state. Vital for \textit{stable} DKS generation is the discovered locking of the soliton to the resonantly enhanced driving pulse. This remarkable mechanism of self-organization adapts the soliton's group velocity by shifting the soliton's central wavelength so that it matches the externally imposed driving pulse repetition rate. As result, the pulse repetition rate as well as the carrier-envelope offset frequency of the soliton are both controlled and stabilized all-optically by the driving pulse train. While by the same mechanism, timing jitter (or phase noise) in the repetition rate of the driving pulse train will imprint itself on the solitons, cavity filtering\cite{beha2015} has the potential of reducing this input noise. Besides the above findings, the presented results also constitute the first observation of temporal dissipative Kerr-cavity solitons in a standing-wave resonator geometry.

Pulsed driving is not only applicable to the microresonator used here, but is transferable to other nonlinear (micro-)resonators, including on-chip, integrated resonators. In conjunction with compact pulsed picosecond lasers or on-chip based continuous wave lasers and modulators, highly integrated and ultra-efficient systems can be envisioned. Such systems would also benefit from the deterministic DKS generation as well as the intrinsic, all-optical control. From a practical perspective it is important to realize that tuning of the driving pulses' carrier-envelope offset frequency can be replaced by tuning of the microresonator e.g. piezo-electrically\cite{papp2013}, electro-optically\cite{jung2014} or by micro-heaters\cite{xue2015}. It is also conceivable to drive microresonators with e.g. standard, low noise femtosecond mode-locked lasers ($f_\mathrm{rep}=0.1\,\mathrm{to}\,1\,\mathrm{GHz}$). While these lasers' repetition rate is much lower than a microresonator's FSR, the high achievable peak power can result in systems whose efficiency surpasses CW driving by far, while at the same time transferring the low noise property of the mode-locked laser to the high-repetition rate soliton.
Conceptually, the presented results bridge the fields of CW driven microresonators and pulse-based non-resonant super-continuum generation and by combining resonant enhancement and pulsed driving, ultra-efficient systems are created. In particular, this enables micro-photonic pulse compressors, ultra-efficient low noise frequency comb sources and resonant supercontinuum generation for applications including optical data transfer and optical spectroscopy. From a scientific perspective the results represent a new way of controlling light in resonant micro-photonic structures and open new horizons for resonant supercontinuum generation and nonlinear photonics driven by temporally and spectrally structured light.

\subparagraph{Funding:}

This work was funded by the SNF grant 200021\_166108 and the Canton of Neuch\^atel. 


\bibliography{arxiv_submission_v2}{}
\bibliographystyle{unsrt}

\end{document}